\documentclass[10pt,letterpaper,twocolumn]{article} %% two column, final layout
\usepackage{ol2}
\usepackage[draft]{hyperref}
\usepackage{amsmath}
\usepackage{graphicx}
\usepackage{floatrow}
\usepackage{epstopdf}
\definecolor{cream}{RGB}{222,217,201}
\graphicspath{{pict/}{./}}
\usepackage[font={small}]{caption}

\usepackage{bm}%

\newcounter{Fig}

\newcommand{\be}{\begin{equation}}
\newcommand{\ee}{\end{equation}}

\begin{document}
\twocolumn[
\title{ Q-factor and absorption enhancement for plasmonic anisotropic nanoparticles}
\author{Wei Liu,$^{1,*}$ Bing Lei,$^{1}$, and Andrey E. Miroshnichenko$^{2}$}
\address{
$^1$College of Optoelectronic Science and Engineering, National University of Defense
Technology, Changsha, China\\
$^2$Nonlinear Physics Centre, Australian National University,
Acton, ACT 0200, Australia\\
$^*$Corresponding author: wei.liu.pku@gmail.com
}
%--------------------------------------------------------------------
\begin{abstract}
We investigate the scattering and absorption properties of anisotropic metal-dielectric core-shell nanoparticles. It is revealed that the radially anisotropic dielectric layer can accelerate the evanescent decay of the localized resonant surface modes, leading to Q-factor and absorption rate enhancement. Moreover, the absorption cross section can be  maximized to reach the single resonance absorption limit. We further show that such artificial anisotropic cladding materials can be realized by isotropic layered structures, which may inspire many applications based on scattering and absorption of plasmonic nanoparticles.
\end{abstract}
\ocis{290.5850, % Scattering, particles
240.6680,   %Surface plasmons,
160.1190.   %Anisotropic optical materials
%350.5500,   %Propagation
} ] %% activate for two-column option
%\maketitle
%---------------------------------------------------------------------
In the field of nanophotonics and many other interdisciplinary subjects, optical waveguides and resonators play an essential role, which provide an indispensable platform for efficient light-matter interactions~\cite{saleh_fundamentals_2013}. Behind different sorts of waveguides and resonators, there are various mechanisms for spatial confinement of light at different scales, of which the total internal reflection (TIR) is probably the most widely employed principle. Nevertheless, many applications (such as high-precision sensing and nanoscale lasing) pose more stringent requirements of stronger electromagnetic localization that convectional TIR based optical devices can barely meet. Recently a modified version of TIR has been introduced (the so called relaxed total internal reflection, RTIR), which shows that at the interface of anisotropic and isotropic dielectrics, evanescent waves could decay much faster~\cite{Jahani2014_Optica_transparent}.  Based on such a more general principle of TIR, stronger electromagnetic field confinement has been demonstrated in both optical waveguides~\cite{Jahani2014_Optica_transparent} and resonators~\cite{liu_q-factor_2016}, which directly leads to all-dielectric waveguiding beyond the diffraction limit and low order resonances of significantly higher Q-factors.

Besides TIR, electromagnetic surface waves~\cite{Boardman1982_book,polo_electromagnetic_2013} provide an alternative significant approach for photonic energy confinement, of which the subject of surface plasmon waveguides and resonators stands as currently the most outstanding example~\cite{Maier2007}. For many applications based on plasmonic structures, such as nano-scale lasing~\cite{Oulton2009_nature,Noginov2009_nature}, sensing~\cite{Kabashin2009_NM}, and photovoltaic devices~\cite{Atwater2010_NM}, efficient manipulation of scattering and absorption for plasmonic resonances are required, which usually involves special structure design and geometric tuning~\cite{feigenbaum_ultrasmall_2008,min2009_nature_high,Miroshnichenko2010_RMP,Ruan2010_PRL,Lukyanchuk2011_NM,Liu2014_arXiv_Geometric}.

%-------------------------------------------------------------------------------
\begin{figure}
\centerline{\fbox{\includegraphics[width=8.8cm]{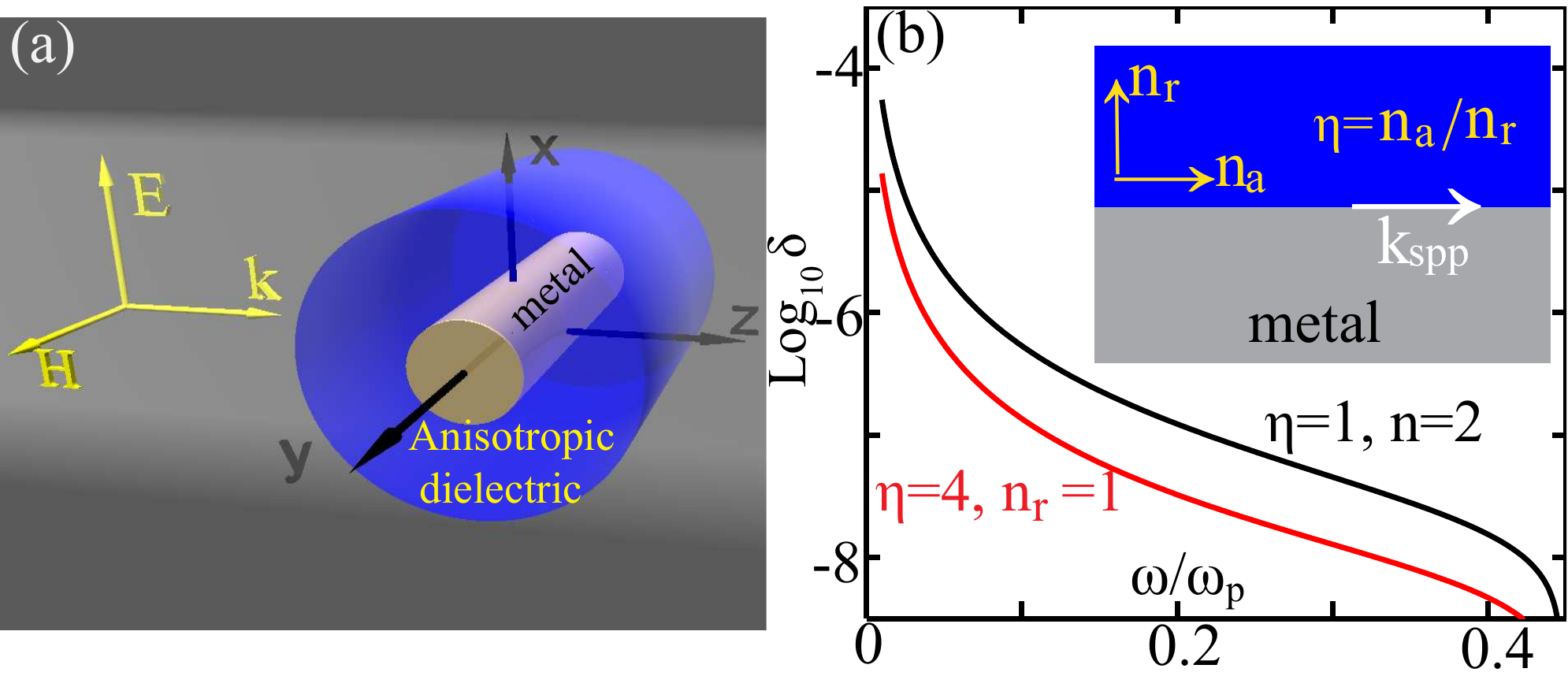}}}\caption{(a) Schematic illustration of the scattering of a normally incident TM polarized plane wave by a metal core- dielectric shell nanowire. The radii for the core and shell layers are $R_1$ and $R_2$ respectively, and the dielectric layer is radially anisotropic on the $x-z$ plane of azimuthal and radial indexes  $n_a$ and $n_r$ respectively. (b) The change of decay length with frequency for the plasmonic mode at the semi-infinite metal-anisotropic dielectric interface (inset).  The indexes of the dielectric layer are $n_a$ and $n_r$  along the propagating and perpendicular directions respectively. The anisotropy parameter is defined as $\eta=n_{a}/n_{r}$.}
\label{fig1}
\end{figure}
%-------------------------------------------------------------------------------

In this Letter, inspired by the principle of RTIR, the effects of anisotropic dielectric layers on the localized surface plasmonic resonances are studied. It is demonstrated that the anisotropy can effectively provide an additional degree of freedom for manipulation of scattering and absorption cross sections. More specifically, we investigate the scattering of a core (metal)-shell (radially anisotropic dielectric) nanowire and show that the dielectric anisotropy accelerates the decay of the plasmonic mode at the metal-dielectric interface, leading to  higher Q-factors of the plasmonic resonances. It is further revealed that, due to stronger light confinement effect induced by the anisotropic layer, the resonance absorption can be maximized to reach the single resonance absorption limit with properly designed resonators.  We expect that the mechanism we have proposed can shed new light not only to many plasmonic structure based applications, such as nano-lasing, sensing, and photovoltaic devices, but also to the investigations into scattering properties of two dimensional (2D) structures that are intrinsically highly anisotropic~\cite{Xia2014_2D}.

In Fig.~\ref{fig1}(a) we show schematically the nonmagnetic ($\mu=1$) core-shell structure under consideration: the isotropic metallic core of radius $R_1$ with permittivity $\varepsilon_m$ coated by a radially anisotropic dielectric layer of radius $R_2$; On the $x-z$ plane the radial and azimuthal refractive indices are $n_r$ and $n_a$ respectively and the anisotropy parameter is defined as $\eta=n_{a}/n_{r}$; the normally incident plane wave is of transverse magnetic (TM) polarization (the magnetic field is fixed along the $y$ direction, axis of the nanowire) to ensure the excitation of the plasmonic modes. We note here that for the transverse electric polarization, as there is only one electric field component along the $y$ direction and thus the mode is not affected by the radial anisotropy on the $x-z$ plane.

To understand the effects of the anisotropic layer on the plasmonic type resonances, we start with the fundamental waveguiding structure of semi-infinite metal-anisotropic dielectric configuration [shown as the inset of Fig.~\ref{fig1}(b)]: the dielectric indices along the propagating and perpendicular directions are set to be $n_a$ and $n_r$ respectively [consistent with the dielectric layer in Fig.~\ref{fig1}(a)]. The angular wavenumber of the supported TM plasmonic mode is~\cite{Boardman1982_book}:

%--------------------------------------------------------------
\begin{equation}
\label{kspp}
k_{\rm spp}  = k_0\sqrt {{{\varepsilon _m \varepsilon _d (\varepsilon _d  - \varepsilon _m )} \over {\varepsilon _d^2  - \eta ^2 \varepsilon _m^2 }}},
\end{equation}
%-----------------------------------------------------------
where $k$ is the angular wave number in the background material (vacuum in this study) and $\varepsilon _d=n _a^2 $. As a result, the angular wave number along the perpendicular direction in the dielectric layer can be expressed as~\cite{saleh_fundamentals_2013,Jahani2014_Optica_transparent}:

%--------------------------------------------------------------
\begin{equation}
\label{k_perpendicular}
k_d^ \bot   = \sqrt {\varepsilon _d k_0^2  - \eta ^2 k_{\rm spp}^2 }
\end{equation}
%-----------------------------------------------------------
Then we define the perpendicular decay length of the plasmonic mode in the dielectric layer as
%--------------------------------------------------------------
\begin{equation}
\label{decay}
\delta  = {\mathop{\rm Im}\nolimits} (k_d^ \bot  )^{ - 1},
\end{equation}
%-----------------------------------------------------------
where $\rm Im(\cdot)$ corresponds to the imaginary part. To demonstrate how the anisotropic dielectric layer accelerates the evanescent decay of the plasmonic mode, in Fig.~\ref{fig1}(b) we show the decay length for both isotropic and anisotropic cases: the  refractive index of isotropic dielectric layer is fixed at $n=2$ and for better comparison the indexes of the anisotropic dielectric layer are restricted by $n_an_r=n^2$, and as a result $\eta=(n_a/n)^2$ (as is the case throughout this paper); without lose of generality, we employ the Drude model for the metal $\varepsilon _m  = 1 - \omega _p^2 /\omega ^2$ where $\omega$ is the angular frequency and $\omega_p$ is the bulk plasmon frequency. It is clear from Fig.~\ref{fig1}(b) that for the anisotropic case, the decay length is much smaller, indicating much faster exponential decay along the perpendicular direction.

%-------------------------------------------------------------------------------

\begin{figure}
\centerline{\fbox{\includegraphics[width=8.8cm]{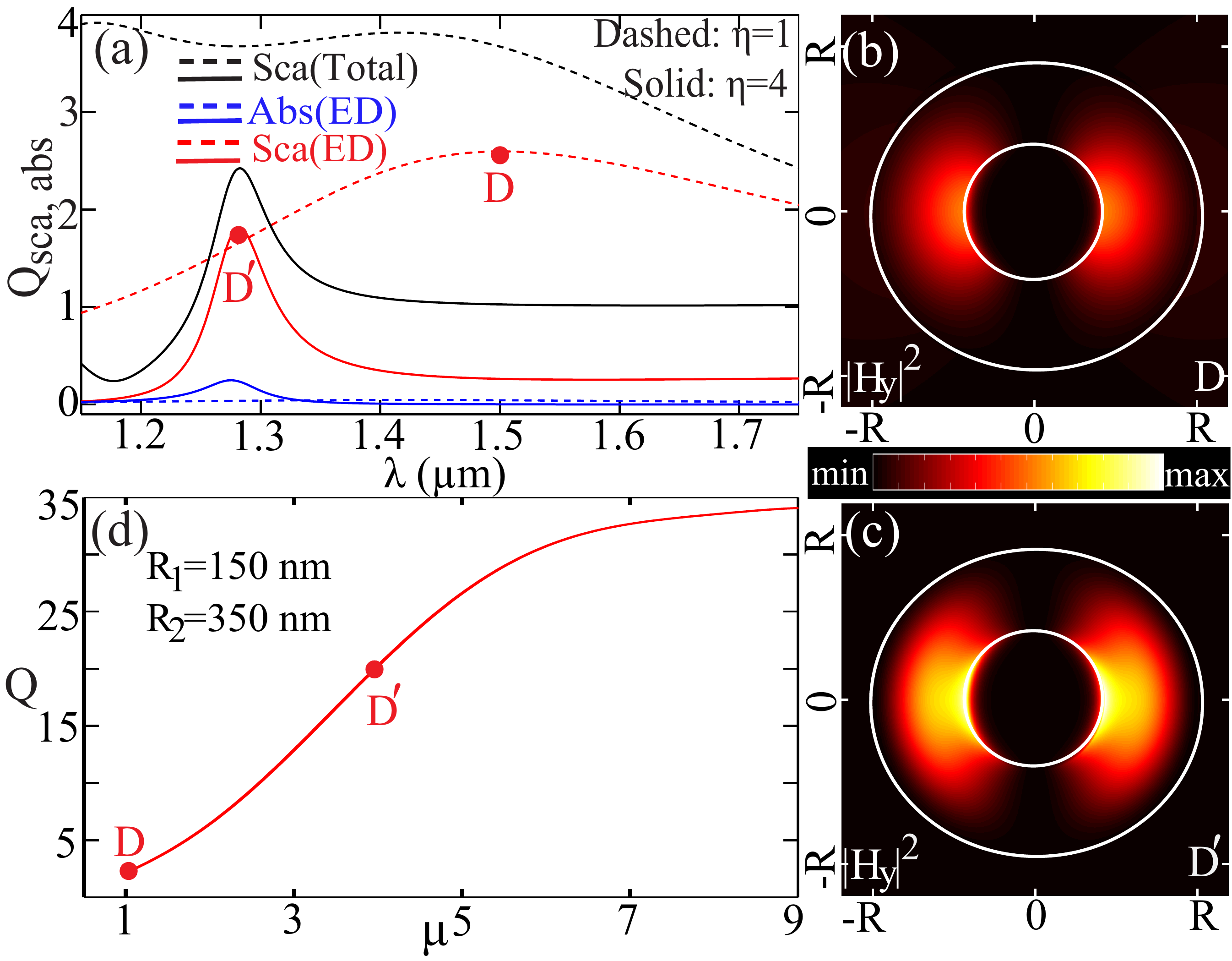}}}\caption{(a) Scattering (red curves) and absorption (blue curves) efficiency spectra (both total efficiency and that of the ED) for the core-shell nanowire of $R_1=150$ nm and $R_2=350$ nm. The results for both the isotropic ($\eta=1$, dashed curves) and anisotropic ($\eta=4$, solid curves) cases are shown. The scattering resonant positions of ED are denoted by $D$ and $D'$ respectively ($\lambda_D=1.5~\mu$m and $\lambda_{D'}=1.28~\mu$m) and the corresponding near-field distributions (in terms of $|H_y|^2$) are shown in (b) and (c) . (d) The dependence of Q-factor of the ED resonance on the anisotropy parameter, where the two cases shown in (a) are also indicated. }
\label{fig2}
\end{figure}
%-------------------------------------------------------------------------------

 According to the correspondence between localized and propagating surface plasmons~\cite{Liu2014_arXiv_Geometric}, similar to the demonstration of Q-factor enhancement for RTIR based all-dielectric resonators~\cite{liu_q-factor_2016}, we expect that the anisotropic dielectric layer can also improve the Q-factor of plasmonic resonances. To justify this point, now we study directly the scattering properties of the configuration shown in Fig.~\ref{fig1}(a). This seminal 2D scattering problem has been well studied~\cite{Kerker1969_book,chen2012_PRA_anomalous,Chen2013_OE_Tunablity,liu_q-factor_2016}, and the scattering and absorption cross sections can be expressed respectively as:
%--------------------------------------------------------------
\begin{eqnarray}
\label{CMT}
\begin{array}{l}
C_{\rm sca} = {4\over {k}}(|a_0|^2+2\sum\nolimits_{m =1}^\infty|a_m|^2),\\\\
C_{\rm abs} = {4\over {k}}[\Upsilon(a_0)+2\sum\nolimits_{m =1}^\infty {\Upsilon(a_m)}]
\end{array}
\end{eqnarray}
%-------------------------------------------------------------
where the function $\Upsilon(x)$ is defined as $\Upsilon(x)=\rm Re(x)-|x|^2$, and $\rm Re(\cdot)$ means the real part; $a_0$ and $a_m$ are the scattering coefficients, which are $\eta$ dependent. Then the scattering and absorption efficiencies are $Q_{\rm sca, abs}=C_{\rm sca, abs}/2R_2$. Here in this work we confine our study to the plasmonic mode of the lowest order (electric dipole, ED), which corresponds to the scattering coefficient $a_1$~\cite{Vynck2009_PRL,Liu2013_OL2621} and it is actually quite natural to extend such study to other higher order plasmon resonances of TM nature.

In Fig.~\ref{fig2}(a) we show the scattering and absorption efficiencies (the total efficiency and that contributed by ED only) for the nanowire of $R_1=150$ nm and $R_2=350$ nm for both isotropic case of $\eta=1$ and anisotropic case of $\eta=4$. The metal is gold with the permittivity taken from experimental data~\cite{Johnson1972_PRB}. It is clear that the anisotropic cladding makes the ED resonance sharper, indicating a higher Q-factor. To clarify the origin we choose the two scattering resonant positions indicated in Fig.~\ref{fig2}(a) ($\lambda_D=1.5~\mu$m and $\lambda_{D'}=1.28~\mu$m) and the corresponding near-field distributions ($|H_y|^2$) at those two points are shown in Fig.~\ref{fig2}(b) and Fig.~\ref{fig2}(c) respectively. As is the case for the anisotropic metal-dielectric waveguide [shown in Fig.~\ref{fig1}(b)], the sharper resonance is induced by the tighter localization of the plasmonic mode at the interface along the radial direction [see Fig.~\ref{fig2}(b) and (c)], which enables stronger electromagnetic energy confinement. To give more details concerning how the Q-factor of the resonance is affected by $\eta$, in Fig.~\ref{fig2}(d) we show specifically its dependence, where the Q-factor is approximated by fitting the scattering curves into a Lorentzian distribution~\cite{saleh_fundamentals_2013,liu_q-factor_2016}. It is quite clear that the anisotropic dielectric layer can effectively enhance the Q-factor of the plasmonic resonances.

%-------------------------------------------------------------------------------
\begin{figure}
\centerline{\fbox{\includegraphics[width=6cm]{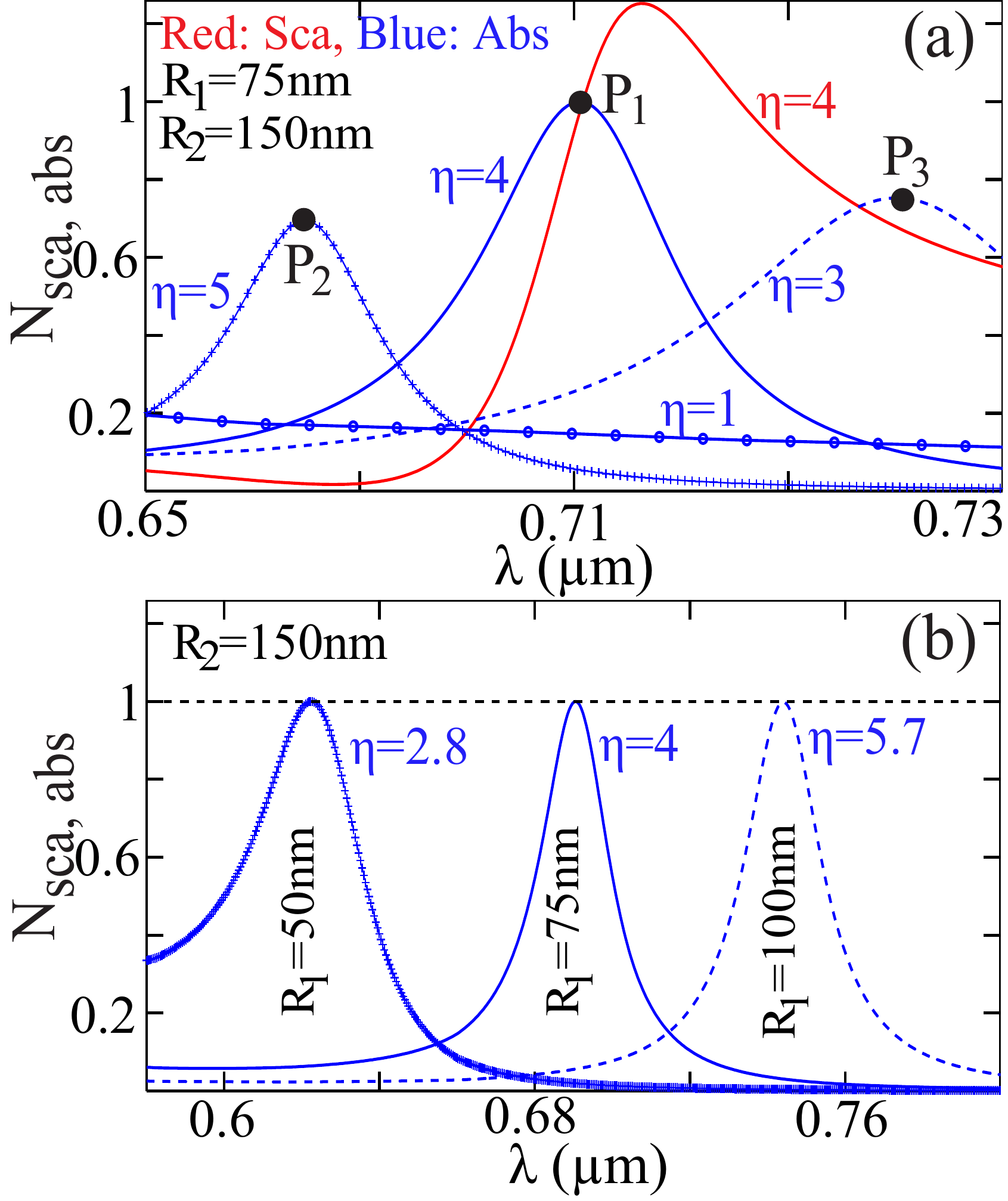}}}\caption{((a) Normalized absorption cross section spectra (blue curves) for the core-shell nanowire of $R_1=75$~nm and $R_2=150$~nm. Four cases of $\eta=1, 3, 4, 5$ are studied and the absorption resonant positions for the anisotropic cases of $\eta\neq1$ are indicated by  $P_{1,2,3}$.  For $\eta=4$ when the single resonance absorption limit can be reached, the normalized scattering cross section spectrum (red curve) is also shown. (b) Normalized absorption cross section spectra for core-shell nanowires of fixed shell radius of $R_2=150$~nm. Three combinations of core radius and anisotropy parameter are shown, where the absorption has been maximized at different resonant positions. }
\label{fig3}
\end{figure}
%-------------------------------------------------------------------------------

In addition to the Q-factor improvements, the results in Fig.~\ref{fig2}(a) also demonstrate that the enhancement of the absorption at the resonance (see blue curves). As a next step, we study in detail how the anisotropy can be employed for the maximization of the plasmonic resonance absorption~\cite{DEPINE_excitation_1995}.  For the 2D scattering configuration shown in Fig.~\ref{fig1}(a), it is known that there exists a single channel absorption cross section limit for each individual resonance~\cite{Seok2011_radiation,Hamam2007_PRA,Ruan2010_PRL,Liu2014_arXiv_Geometric,miroshnichenko_ultimate_2016}:

%--------------------------------------------------------------
\begin{equation}
\label{limit}
C_{\rm lim}=\frac{\lambda}{4 \pi},
\end{equation}
%-----------------------------------------------------------
 which can be reached if at the resonant position the scattering and absorption rates equal to each other~\cite{Liu2014_arXiv_Geometric,Fleury2014_PRB,Tretyakov2013_arXiv,miroshnichenko_ultimate_2016}.  Despite the magnetic dipole resonance (characterized by $a_0$), all other resonances including ED actually correspond two degenerate channels~\cite{Vynck2009_PRL,Liu2013_OL2621}, of which the absorption limit is consequently  $2C_{\rm lim}$. In Fig.~\ref{fig3}(a) we show the normalized ED absorption cross sections $N_{\rm abs}=C_{\rm abs}/2C_{\rm lim}$ for the core-shell nanowire of $R_1=75$~nm and $R_2=150$~nm. Four cases are investigated for $\eta=1, 3, 4, 5$.  It is clear that in the spectrum region shown, the absorption has been significantly enhanced when the anisotropic shell is introduced. For the three anisotropic cases the absorption resonant positions are indicated, where at $P_1$ of $\eta=4$ the single resonance absorption limit is reached. For this special case, we also show in Fig.~\ref{fig3}(a) the normalized ED scattering spectra $N_{\rm sca}=C_{\rm sca}/2C_{\rm lim}$. It is clear that at $P_1$ the scattering and absorption rates (cross sections) are equal, which leads to the absorption maximization. At the other two points $P_2$ ($\eta=5$) and $P_3$ ($\eta=3$), the scattering rates are smaller and larger, respectively, than the absorption rates, leaving the single resonance absorption limit  untouched. Moreover, a proper geometric tuning and selection of the anisotropy parameters can help to achieve the optimal absorption at various resonant frequencies. In Fig.~\ref{fig3}(b) we show three cases with fixed shell layer width of $R_2=150$~nm while with different combinations of core radii and anisotropy parameters, when the absorption reaches the maximum value at different wavelengths. We note that here we demonstrate the absorption maximization in the visible spectral region where the loss of Au is relatively highly. Such enhancement can certainly be achieved at other spectrum regions of low intrinsic loss rate, where however larger anisotropy parameters are required.

\begin{figure}
\centerline{\fbox{\includegraphics[width=8.8cm]{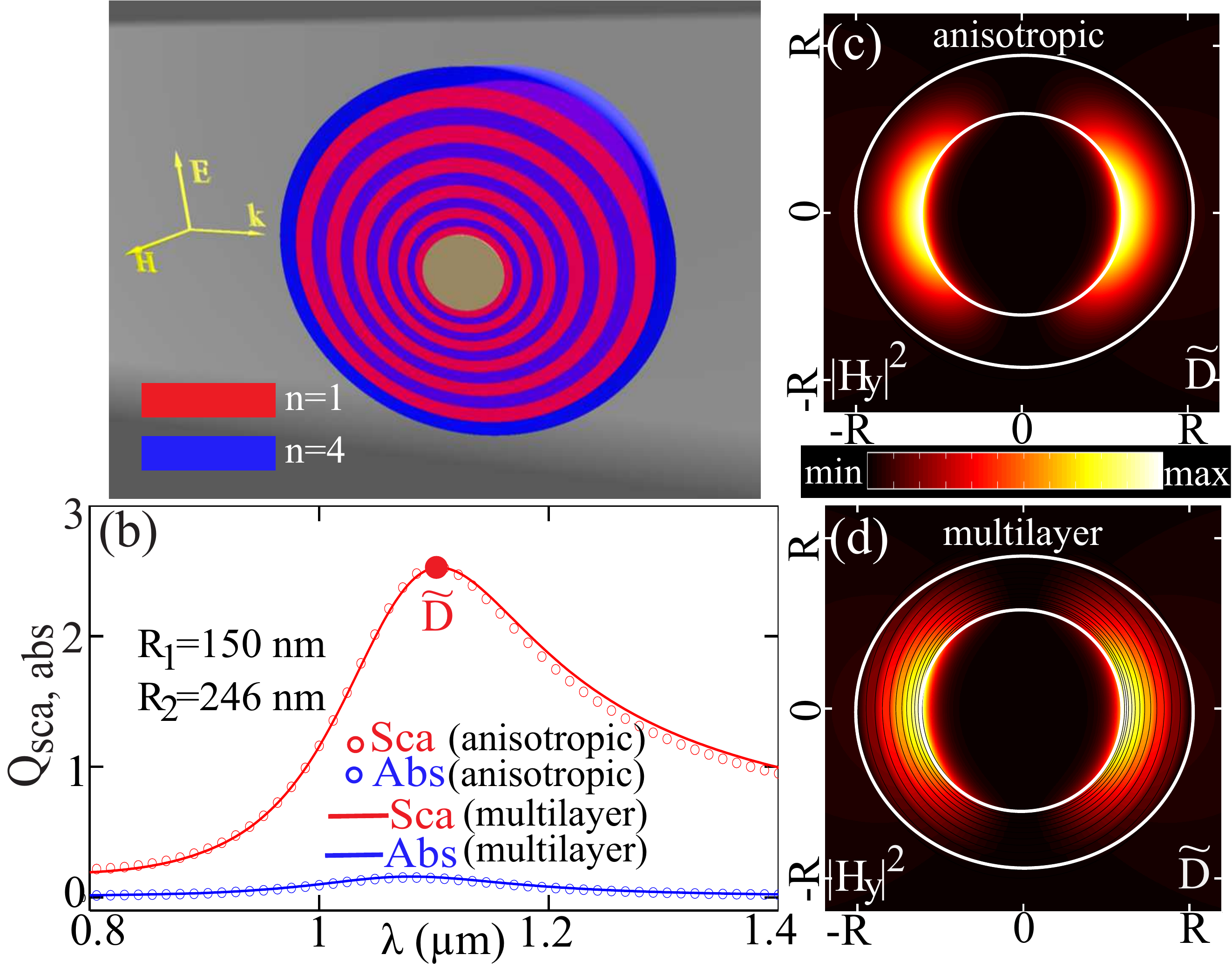}}}\caption{(a) The schematic of the scattering configuration where a relatively large effective anisotropy parameter can be obtained for the cladding layer: the metallic core $R_1=150$~nm is surrounded by alternating isotropic dielectric layers of indexes of $n_1$ and $n_2$ with linearly increasing layer widths ($R_2=246$~nm). The corresponding  scattering and absorption efficiency spectra are shown in (b) by solid curves, where the spectra for the corresponding two layered metal-anisotropic dielectric nanowire are also shown by circles for comparison. The resonant position of $\lambda=1.1~\mu$m is indicated and at this position the near-field distributions (in terms of $|H_y|^2$) are shown for both cases in (c) and (d).}
\label{fig4}
\end{figure}

 Finally, we discuss the possible realizations of the scattering resonators discussed above. For natural materials, the anisotropy parameters are usually low and thus would not significantly affect the Q-factor or the absorption of the low order plasmonic resonances. Nevertheless, with the recent rapid progress in the field of metamaterials, many exotic features of materials, including unusual refractive indexes and anisotropy parameters have been obtained~\cite{choi2011_nature_terahertz,poddubny2013_NP_hyperbolic,wu2014_PRX_electrodynamical}. To achieve large anisotropy parameters of the dielectric layer, we employ a multilayered cladding consisting of two isotropic dielectrics of positive refractive indexes $n_1$ and $n_2$ [$n_1< n_2$, see Fig.~\ref{fig4}(a)].  As long as each dielectric layer width is far smaller than the effective wavelength in the layers, the effective medium theory can be applied~\cite{liu_q-factor_2016,poddubny2013_NP_hyperbolic}, then the effective indexes for the multi-layered cladding along the radial and azimuthal directions are respectively:

\begin{equation}
\label{effective1}
n_{\rm {r}} =n_1n_2/\sqrt {(1 - f)n_1^2  + fn_2^2}, n_{\rm {a}} =\sqrt {fn_1^2  + (1 - f)n_2^2 }, %
\end{equation}
where $f$ is the filling factor of the higher index layer in terms of layer width. As a result, the radial anisotropy parameter obtained is:
%--------------------------------------------------------------
\begin{equation}
\label{effective_medium}
\eta(f)=  \frac{\sqrt {fn_1^2  + (1 - f)n_2^2 } \sqrt {(1 - f)n_1^2  + fn_2^2 }}{n_1n_2}.
\end{equation}
%-----------------------------------------------------------
 The highest anisotropy parameter obtainable is  $\eta_{\rm max}=(n_1^2  + n_2^2 )/2n_1n_2$ when $f=0.5$.  To conform to the index parameters adopted already in this work, we study the simple case of $n_1=1$ and $n_2=4$ which makes effectively $n_an_r=n^2=4$ and $\eta_{\rm max}=2.13$ when $f=0.5$. To achieve  $\eta_{\rm max}$, as long as the effective medium theory can be applied and $f=0.5$, the specific layer width is not important and thus of course the whole cladding does not have to bee periodic. We give a simple example here: the metal core ($R_1=150$~nm is covered by six dielectric unit cells and each unit cell is made of two dielectric layers with different indexes but the same width (this makes $f=0.5$); each layer width in the $i-th$ unit cell is $d_i=2i+1$ for $i=1:6$ [see the schematic shown in Fig.~\ref{fig4}(a)] and consequently $R_2=246$~nm. The scattering and absorption efficiency spectra of such isotropic multi-layered resonator are shown in Fig.~\ref{fig4}(b) (solid curves), which agree very well with the results (shown in circles) of the corresponding two-layered metal-anisotropic dielectric nanowire of $n_an_r=n^2=4$ and $\eta_{\rm max}=2.13$. Furthermore, we show also the near-field distributions of $|H_y|^2$ at the scattering resonant position [indicated by $\widetilde{D}$ in Fig.~\ref{fig4}(b) and $\lambda_{\widetilde{D}}=1.1~\mu$m] in Fig.~\ref{fig4}(c) and Fig.~\ref{fig4}(d) for both cases, which also agree quite well. Those results can justify the effectiveness of our approach for resonance control relying on dielectric anisotropy.

To conclude, we propose and demonstrate the efficient engineering of Q-factor and absorption for plasmonic resonances relaying an anisotropic cladding layer. By analyzing the scattering properties of the anisotropic core-shell metal-dielectric nanowires,  we reveal that the resonance Q-factor can be enhanced and the absorption can be maximized to reach the ultimate single resonance limit due to the faster evanesce decay and thus stronger energy confinement of the localized surface plasmonic  modes. We note that the principle we have revealed can certainly be applied to higher order modes, and resonators of other shapes, and also  other anisotropy such as permeability anisotropy, and to other sorts of surface modes~\cite{polo_electromagnetic_2013}. For the proof of concept demonstration,  we adopt a multi-layered lossless cladding that exhibits moderate effective anisotropy parameter ($\eta_{\rm max}=2.13$). Nevertheless, we have to keep in mind that relying on metamaterials and 2D materials,  much more exotic anisotropy features can be obtained (including complex anisotropy parameters)~\cite{choi2011_nature_terahertz,poddubny2013_NP_hyperbolic,wu2014_PRX_electrodynamical,Xia2014_2D}, which may enable much stronger field localization and more flexible resonance manipulations, thus shedding new light to many plasmonic metamateris and 2D materials based fundamental researches and applications.

%\section*{Acknowledgement}
We thank D. A. Powell, D. N. Neshev and Yuri S. Kivshar for useful discussions, and acknowledge the financial support from the National Natural Science Foundation of China (Grant number: $11404403$). W. L. thanks the Nonlinear Physics Centre for a warm hospitality during his visit to Canberra.

%%\bibliographystyle{ol}
%%\bibliography{References_scattering}
%\bibliographystyle{ol}
%\bibliography{References_scattering}
%\newpage
%\bibliographystyle{osajnlnt}
%\bibliographyfullrefs{References_scattering}

%==========================
\end{document}